\documentclass[twocolumn,superscriptaddress,amssymb, nobibnotes, aps, prb]{revtex4-2}
\usepackage{enumitem}
\usepackage{romannum}
\usepackage{amsthm}
\usepackage{hhline}
\usepackage{epsfig,amsopn}
\usepackage{graphicx}
\usepackage{xcolor}
\usepackage[english]{babel}
\usepackage{amsmath,amssymb}
\usepackage[colorlinks,allcolors=blue]{hyperref}
\usepackage{url}
\usepackage{multirow}
\usepackage{enumerate}
\usepackage{verbatim}
\usepackage{bm}
\begin{document}
\title{Anomalous dynamical response of non-Hermitian topological phases}
\author{Ritu Nehra}
\affiliation{Raman Research Institute, Bangalore-560080, India}
\affiliation{Department of Physics, Ben-Gurion University of the Negev, Beer-Sheva 84105, Israel}
\email{ritu@post.bgu.ac.il}
\author{Dibyendu Roy}
\email{droy@rri.res.in}
\affiliation{Raman Research Institute, Bangalore-560080, India}
\begin{abstract}
Composite topological phases with intriguing topology like M{\"o}bius strips emerge in sublattice symmetric non-Hermitian systems due to spontaneous breaking of time-reversal symmetry at some parameter regime. While these phases have been characterized by nonadiabatic complex geometric phases of multiple participating complex bands, the physical properties of these phases largely remain unknown. We explore the dynamical response of these phases by studying Loschmidt echo from an initial state of the Hermitian Su-Schrieffer-Heeger (SSH) model, which is evolved by a non-Hermitian SSH Hamiltonian after a sudden quench in parameters. Topology-changing quenches display non-analytical temporal behavior of return rates (logarithm of the Loschmidt echo) for the non-Hermitian SSH Hamiltonian in the trivial,  M{\"o}bius and topological phase. Moreover, the dynamical topological order parameter appears only at one side of the Brillouin zone for the M{\"o}bius phase case in contrast to both sides of the Brillouin zone for quench by the trivial and topological phase of the non-Hermitian SSH model. The last feature is a dynamical signature of different symmetry constraints on the real and imaginary parts of the complex bands in the M{\"o}bius phase.  
\end{abstract}
\maketitle
\section{Introduction}
Topology and quantum dynamics are shown to be intrinsically related in the early research on topology in condensed matter physics \cite{thouless1998topological}. Several near-equilibrium dynamical quantities like linear electrical transport \cite{Thouless1980,Thouless1983}, change in electrical polarization \cite{KingSmith1993}, and Josephson current \cite{Kitaev2001} are directly connected to the underlying topology of the band structures. Recent studies have further extended the realm of their connections to the out-of-equilibrium regime. It has been found that topology-changing quenches are always followed by non-analytical temporal behavior of return rates (logarithm of the Loschmidt echo), characterizing dynamical phase transitions (DPTs)~\citep{PhysRevB.93.085416,PhysRevB.91.155127,PhysRevB.100.224307}. Topological edge modes can further influence features of non-equilibrium transport in open topological systems \cite{Roy2013,Bondyopadhaya2019}. The effective Hamiltonian of an open quantum system is non-Hermitian. In recent years, there has been massive research interest in exploring topological features in non-Hermitian models \citep{vyas_topological_2021, ritutopo2022, PhysRevB.107.085426, jin_topological_2017, longhi_non-hermitian_2015, ghatak_observation_2020, jiang_multiband_2021, berry_physics_2004, cartarius_exceptional_2009, xu_topological_2016}.

The energies are complex-valued in non-Hermitian systems in contrast to the Hermitian systems with real eigenvalues~\citep{hatano_localization_1996, okugawa_non-hermitian_2021, wang_topological_2021, hu_knots_2021, PhysRevLett.124.056802}. The degeneracy in complex energy spectrum leads to the non-analyticities (singularities) coined as exceptional points (EPs)~\citep{heiss_physics_2012, berry_physics_2004, heiss_chirality_2008}. The complex energy spectrum and the EPs pave the way for the ramification and unification of various crucial symmetries in non-Hermitian systems~\citep{kawabata_symmetry_2019,kawabata_topological_2019,kawabata_parity-time-symmetric_2018}. Such unification and branching out of symmetries have led to intriguing topological phases, both with and without their Hermitian counterparts. For instance, the trivial and topological phase with a complex-energy gap in a sublattice symmetric non-Hermitian Su-Schrieffer-Heeger (NSSH) model~\citep{Natureexp, lieu_topological_2018, yin_geometrical_2018, shen_topological_2018} are similar to those phases with a real-energy gap in the related Hermitian model. The NSSH model also hosts a gapless composite topological phase named the M\"{o}bius phase in the parameter region between two EPs. The M\"{o}bius phase involves the multiple participating complex bands \citep{kawabata_parity-time-symmetric_2018, vyas_topological_2021, ritutopo2022} and does not have a Hermitian analog.

\begin{figure}[h!]
  \includegraphics[width=\linewidth, height=0.6\linewidth]{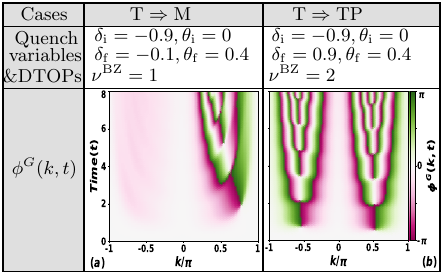}
  \caption{Summary of main results for two quench protocols from a trivial (T) phase of Hermitian SSH model to a M{\"o}bius (M) and a topological (TP) phase of non-Hermitian SSH model. Pancharatnam phase $\phi^G(k,t)$ is calculated for single particle with momentum $k$ at time $t$. Parameters of initial and final Hamiltonian are denoted by $\delta_\mathrm{i}$, $\theta_\mathrm{i}$ and $\delta_\mathrm{f},\;\theta_\mathrm{f}$, respectively. Number of DTOPs per Brillouin zone is represented by $\nu^{\text{BZ}}$, which shows different $\nu^{\text{BZ}}$ for the final Hamiltonian in the M and TP phase.}
\label{fig:intro}
\end{figure}

Our main aim of this study is to identify physical signatures of these non-Hermitian phases particularly the gapless M\"{o}bius phase. We are not aware of any study exploring equilibrium or out-of-equilibrium physical properties of the M\"{o}bius phase yet. We here explore the global dynamical features arising from the integration or disintegration of different crucial symmetries and the accompanying topological phases of the NSSH model. We study Loschmidt echo from an initial state of the Hermitian SSH model, which is evolved by an NSSH Hamiltonian after a sudden quench in parameters. Previous studies on Loschmidt echo with non-Hermitian topological models have mostly considered parity-time ($\mathcal{PT}$) symmetric models~\citep{QIU2019392, Tang_2022, longhi2019loschmidt} apart from few exceptions \cite{Zhou_2021,Mondal2022}. Nevertheless, pertinent questions remain unanswered, particularly for sublattice symmetric models, as shown here. We first show that a quench by the NSSH Hamiltonian in the M\"{o}bius phase exhibits DPTs from both the trivial and topological phase. The last feature is sharply different from the quench by a $\mathcal{PT}$ symmetric SSH Hamiltonian in the broken phase~\citep{QIU2019392,Tang_2022,longhi2019loschmidt} from both the phases when there is no DPT. Further, we also observe unique dynamical signatures related to the symmetry constraints in the composite phase of the sublattice symmetric NSSH model. We demonstrate these unique features in the return rates and the dynamical topological order parameter (DTOP). We here observe two DTOPs in the Brillouin zone (BZ) for a topology changing quench by the NSSH model in either trivial or topological phase, and the initial state in any band. However, there is only one DTOP in the positive or negative half of the BZ when the NSSH model in the  M\"{o}bius phase.  Analyzing the consequences of symmetries on the complex energy spectrum, we explain the differences in the number of the DTOPs for quench by NSSH in the topological or trivial phase versus M{\"o}bius phase. The Fig.~\ref{fig:intro} summarizes our main findings.

The rest of the paper is divided into three sections. In Sec.~\ref{sec1}, we introduce the NSSH model, its complex Bloch vector and spectrum, important symmetries and their consequences on various topological phases. We also briefly describe the non-Hermitian analogs of various dynamical quantities like Loschmidt echo, return rate, and DTOPs, which we apply to analyze DPTs in different quench settings. We give our main results from the various quench protocols in Sec.~\ref{results} and explain these results there. We conclude the paper with a summary of our main findings and some outlook for possible issues and future extensions of this research in Sec.~\ref{conclusion}.

\section{\label{sec1}Model $\&$ quench Dynamics}
\subsection{\label{model}Non-Hermitian SSH model}
The Hamiltonian of the bi-partite sublattice symmetric NSSH chain \citep{ritutopo2022, vyas_topological_2021} reads as 
\begin{align}
H=\displaystyle\sum_{m=1}^{L}(v_\ell a^{\dagger}_{m}b_{m}+v_r b^{\dagger}_{m}a_{m}+w_\ell b^{\dagger}_ma_{m+1}+w_r a^\dagger_{m+1}b_{m}),
\end{align}
where $c^{\dagger}_m\;(c_m)$ represents the spinless fermionic creation (annihilation) operator at $c=a,b$ sublattice site of the $m^{th}$ unit cell, and $L$ is the length of the chain. We consider periodic boundary condition (PBC), e.g., $L+i\equiv i$. Here, $v_\lambda\;(w_\lambda)$ represents the intra-cell (inter-cell) hopping amplitude with a subscript $\lambda=\ell,r$ indicating the left and right direction of hopping, respectively. The PBC allows us to write $H$ as a $2\times 2$ matrix in the momentum $k$ space as 
\begin{align}\label{ham_mmtm}
\mathcal{H}(k)=\begin{bmatrix}
&0&(v_\ell+w_re^{ik}) \\
&(v_r+w_\ell e^{-ik}) &0
\end{bmatrix}.
\end{align}
In the rest of the paper, we parameterize as $v_\ell=v_r=J(1-\delta),\;w_r=w_\ell e^{-\theta}=J(1+\delta),\;\theta>0$, which facilitate symmetric quenches and a single control of non-Hermiticity ($\theta$). The boundaries of various topological phases can be controlled by tuning the system parameters, e.g., $\delta$, $\theta$. The  trivial, M{\"o}bius and topological phase appear, respectively, for $\frac{1-e^{\theta}}{1+e^{\theta}}>\delta$,  $\frac{1-e^{\theta}}{1+e^{\theta}}<\delta<0$ and $\delta>0$ \citep{ritutopo2022, vyas_topological_2021}.

\begin{figure*}
\centering
\hspace{40pt}\textbf{Trivial(T)}\hspace{120pt}\textbf{ M\"{o}bius(M)}\hspace{110pt}\textbf{ Topological(TP)}\\
\includegraphics[width=0.325\linewidth, height=0.2\linewidth]{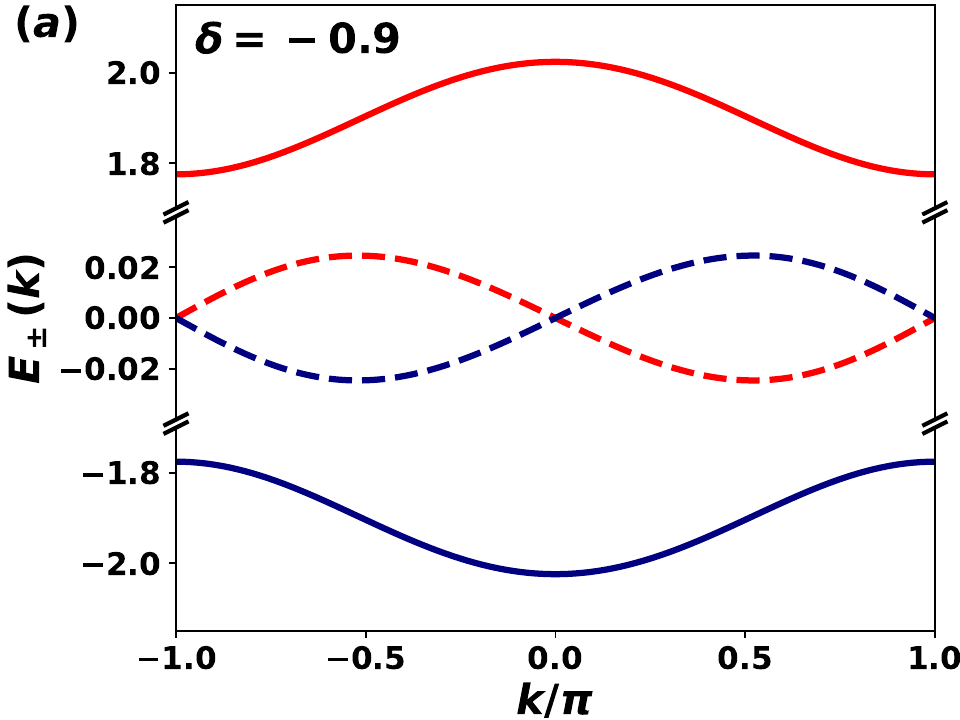}
\includegraphics[width=0.33\linewidth, height=0.2\linewidth]{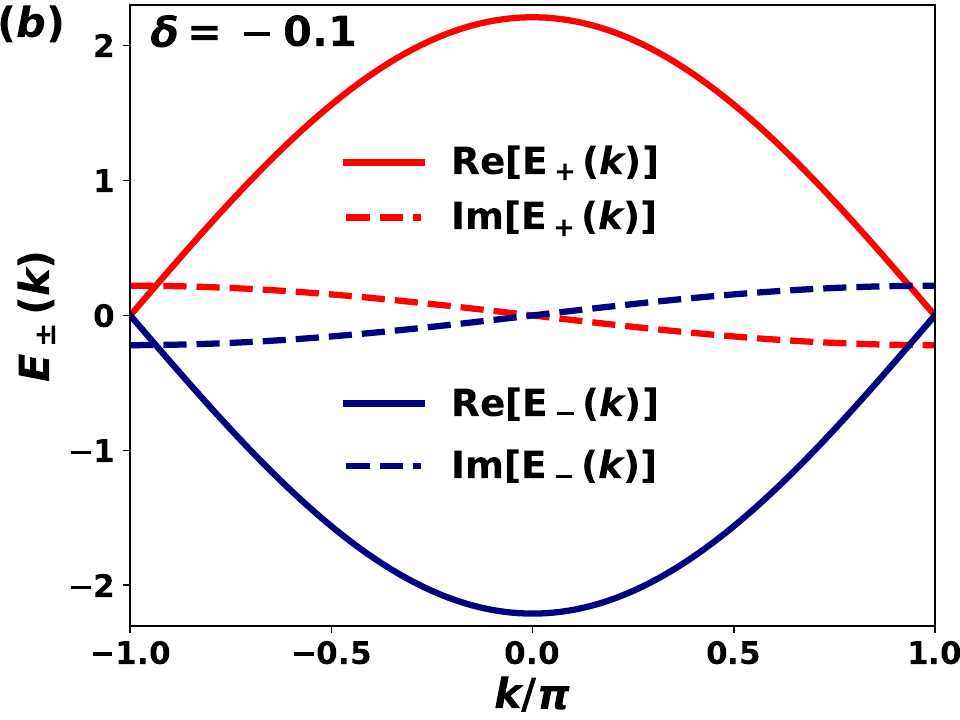}
\includegraphics[width=0.325\linewidth, height=0.2\linewidth]{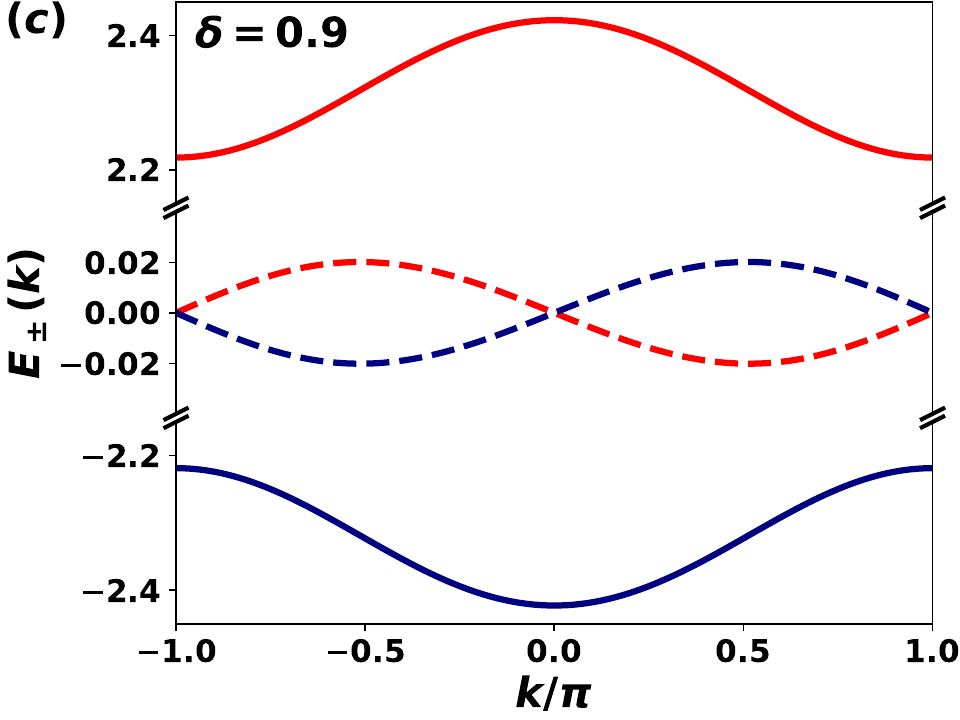}\\
\includegraphics[width=0.325\linewidth, height=0.2\linewidth]{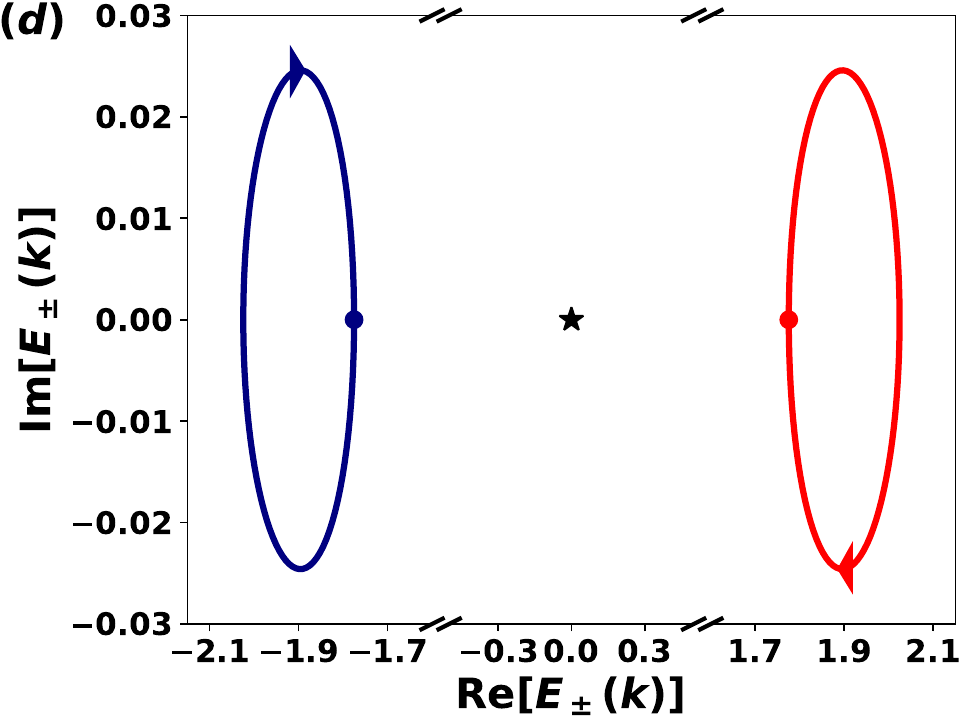}
\includegraphics[width=0.33\linewidth, height=0.2\linewidth]{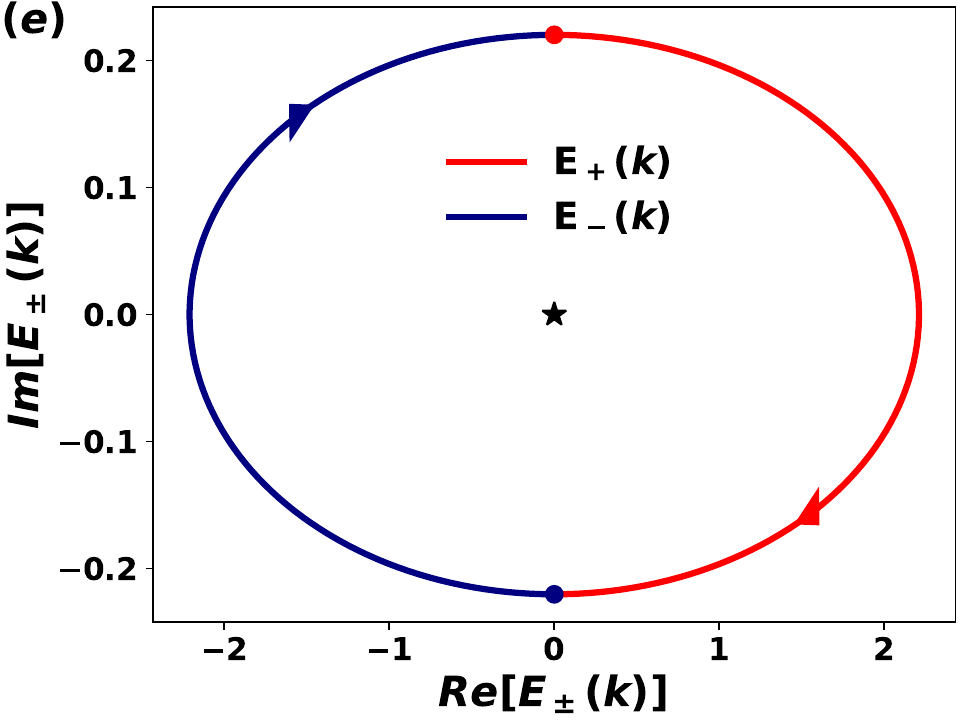}
\includegraphics[width=0.325\linewidth, height=0.2\linewidth]{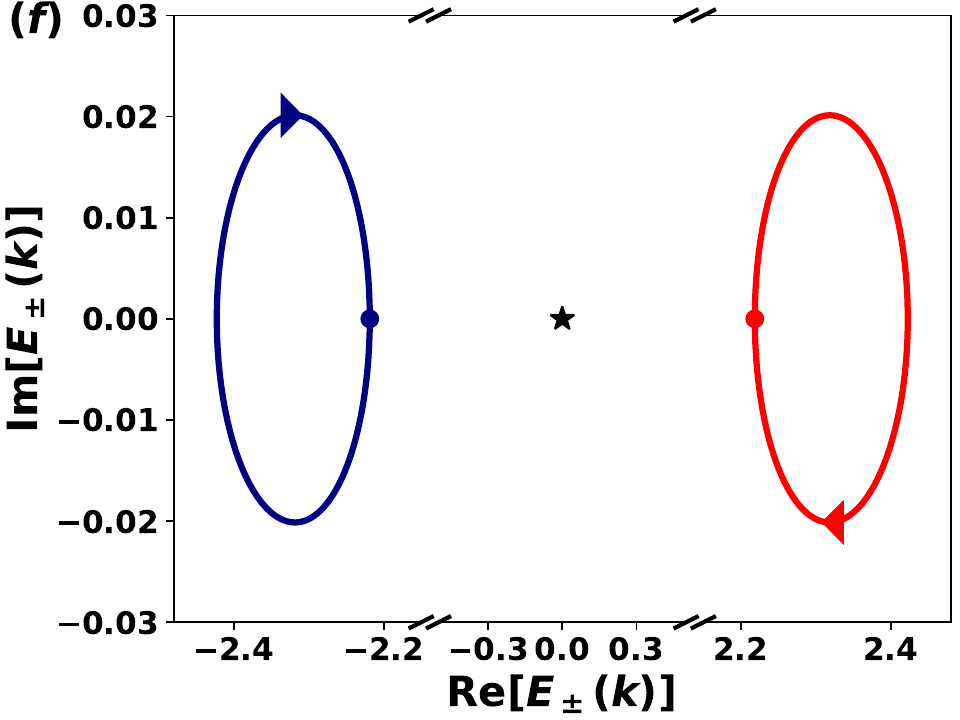}\\
\includegraphics[width=0.325\linewidth, height=0.2\linewidth]{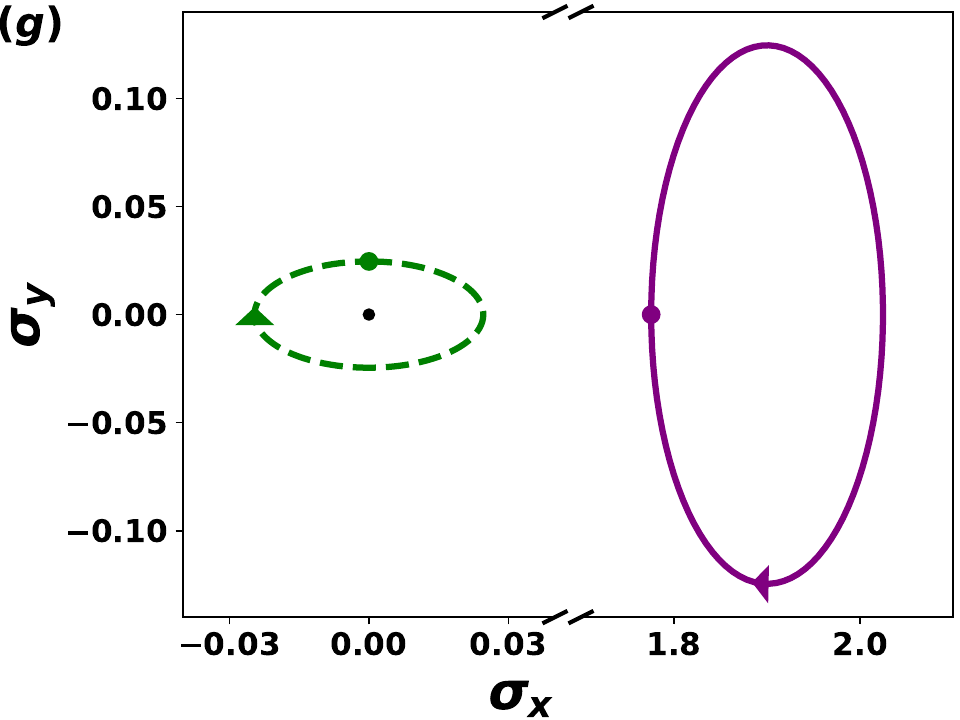}
\includegraphics[width=0.33\linewidth, height=0.2\linewidth]{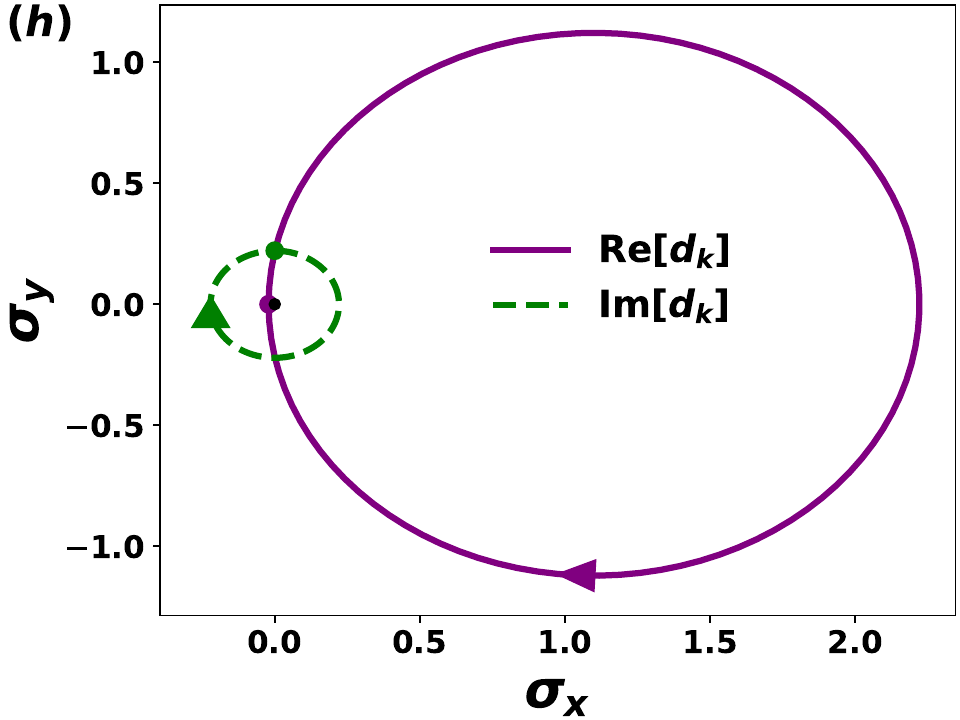}
\includegraphics[width=0.325\linewidth, height=0.2\linewidth]{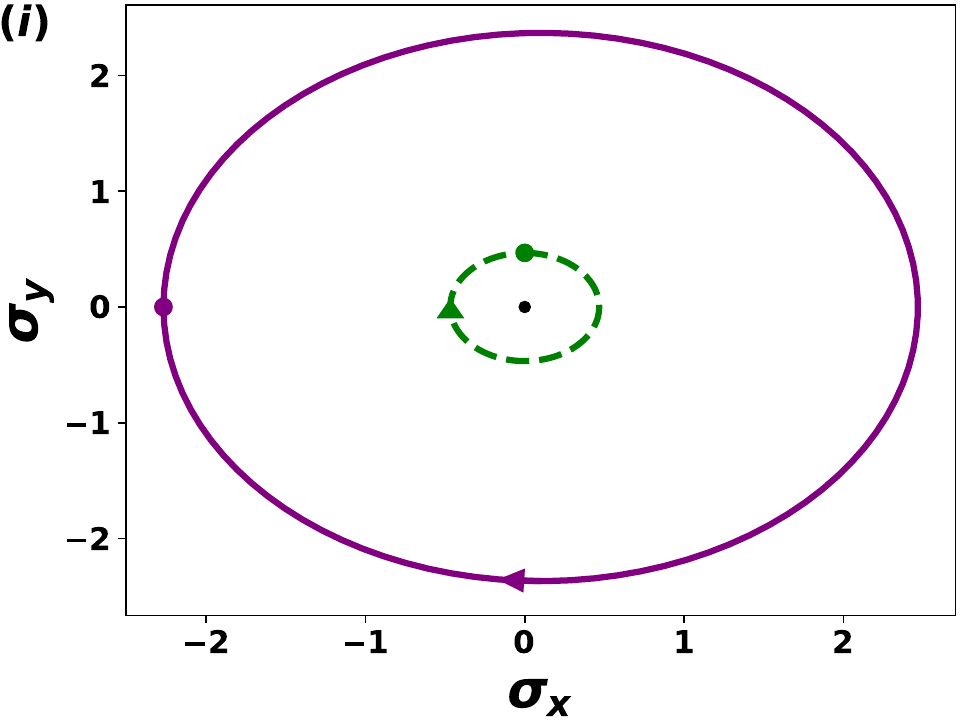}
\caption{The complex energy spectrum $E_{\pm}(k)$ with momentum $k$ (top row) and on the parametric space of ${\text Re}[E_{\pm}(k)]$ and ${\text Im}[E_{\pm}(k)]$ (middle row) in three different phases of non-Hermitian SSH model. The contour of the endpoints of real (full line) and imaginary (dashed line) part of $\vec{d}_k$ on the $d_k^x,d_k^y$ plane as $k$ is swept across the Brillouin zone, $k=-\pi \to \pi$  (bottom row). The parameters are $J=1,\theta=0.4$, and $\delta=-0.9$ (trivial), $-0.1$ (M{\"o}bius), and $0.9$ (topological). The dot represents value at $k=-\pi$ and arrow indicates $k=-\pi \to \pi$ for complex energy $E_\pm(k)$ and Bloch vector $\vec{d}_k$ in the middle and bottom rows. The exceptional point and the origin of $d_k^x,d_k^y$ plane are shown by black star and dot, respectively.}
\label{NSSH}
\end{figure*}

We can further present $\mathcal{H}(k)$ in terms of the Pauli matrices, $\vec{\sigma}=(\sigma_x,\sigma_y,\sigma_z)$, as $\mathcal{H}(k)=\vec{d}_k.\vec{\sigma}$, where  $\vec{d}_k=(d_k^x,d_k^y,d_k^z)\in \mathrm{C}^3$  is a complex-valued three-dimensional Bloch vector with components as 
\begin{align}
d^x_k=J(1-\delta)+J(1+\delta)e^{\frac{\theta}{2}}\cos\Big(k+i\frac{\theta}{2}\Big),\nonumber\\d^y_k=-J(1+\delta)e^{\frac{\theta}{2}}\sin\Big(k+i\frac{\theta}{2}\Big),\;d^z_k=0. \label{bloch}
\end{align}
In the Hermitian limit ($\theta=0$), $\vec{d}_k$ becomes the real-valued Bloch vector $(d^{x,y,z}_k \in \mathrm{R})$ of the SSH model~\citep{su_soliton_1980, asboth_su-schrieffer-heeger_2016}, whose endpoint traces out a closed loop on the $d^x_k,d^y_k$ plane either encircling or excluding the origin as $k$ is swept across the BZ, $k=-\pi \to \pi$. The bulk winding number indicating the topology of the model counts the number of times the loop winds around the origin. In Figs.~\ref{NSSH}(g-i), we show the analogous plots for the real and imaginary parts of $\vec{d}_k$ of the NSSH model for three different values of $\delta~(=-0.9,-0.1,0.9)$. The loop by the imaginary part of $\vec{d}_k$ always includes the origin of the $d^x_k,d^y_k$ plane. However, the loop by the real part of $\vec{d}_k$ shows interesting features in three different phases of the NSSH model. In the topological phase, it winds around both the origin and the loop by the imaginary part without intersecting the loop in Fig.~\ref{NSSH}(i). The loop by the real part of $\vec{d}_k$ avoids the origin as well as the loop by the imaginary part of $\vec{d}_k$ in the trivial phase as shown in Fig.~\ref{NSSH}(g). For the M{\"o}bius phase, the loops by the real and imaginary parts of $\vec{d}_k$ intersect, and the loop by the real part may or may not include the origin like in Fig.~\ref{NSSH}(h). Comparing the Bloch vector between the Hermitian and non-Hermitian case in Eq.~\ref{bloch}, we find a modification of the inter-cell hopping as $J(1+\delta)e^{\frac{\theta}{2}}$, and an emergence of complex-valued momentum $k'=k+i\frac{\theta}{2}$ in the non-Hermitian model, which lead to various kinds of skin effects ~\citep{okuma_topological_2020,yuce_non-hermitian_2020,zhang_tidal_2021,li_critical_2020} in such model.

The complex energy eigenvalues of the two bands of the NSSH model are
\begin{align}
E_{\pm}(k)=\pm\epsilon_k=\pm\sqrt{(v_\ell+w_re^{ik})(v_r+w_\ell e^{-ik})}.
\end{align} 
We show the real (full lines) and imaginary (dashed lines) parts of $E_p(k)$ for $p=\pm$ with $k$ in Figs.~\ref{NSSH}(a-c) for three different values of $\delta~(=-0.9,-0.1,0.9)$ representing the trivial, M{\"o}bius and topological phases of the non-Hermitian chain.  The bi-orthogonal eigenvectors to the corresponding eigenvalues $\pm\epsilon_k$ are
 \begin{align}
 |\psi^\pm_k\rangle=\frac{1}{\sqrt{2}}\Big(
 \frac{v_\ell+w_re^{ik}}{\epsilon_k}\tilde{a}^\dagger_{k}\pm \tilde{b}^\dagger_{k} \Big)|0\rangle,\\
 \;\langle\chi^\pm_k|=\langle 0|\frac{1}{\sqrt{2}}\Big(
 \frac{v_r+w_\ell e^{-ik}}{\epsilon_k}\tilde{a}_{k}\pm \tilde{b}_{k} \Big),
 \end{align}
 where $\tilde{c}^\dagger_{k}(\tilde{c}_{k})$ is the fermionic creation (annihilation) operator at $c=a, b$ sublattice with a momentum $k$.

 The NSSH chain with non-reciprocal hoppings has anti-unitary symmetries, time-reversal symmetry (TRS) $\mathcal{T}K$ and particle-hole symmetry (PHS$^{\dagger}$) $\mathcal{C}K$, and unitary sublattice symmetry (SLS) $\mathcal{S}$ given by the following relations: $\mathcal{T}^{-1}\mathcal{H}^*(k)\mathcal{T}=\mathcal{H}(-k)$, $\mathcal{C}^{-1}\mathcal{H}^*(k)\mathcal{C}=-\mathcal{H}(-k)$ and $\mathcal{S}^{-1}\mathcal{H}(k)\mathcal{S}=-\mathcal{H}(k)$ where $\mathcal{T}=\sigma_0, \mathcal{C}=\sigma_z,\mathcal{S}=\sigma_z$ with $\sigma_0$ being $2 \times 2$ identity matrix and $K$ is the complex-conjugate operator~\citep{andp.202300133}. The complex energy bands individually respect the TRS in all three phases: $E_{\pm}(k)=E_{\pm}^*(-k)$. The SLS indicates the bands appear as an opposite-sign pairs. The TRS gives, Re$[E_{\pm}(k)]=$Re$[E_{\pm}(-k)]$, and Im$[E_{\pm}(k)]=-$Im$[E_{\pm}(-k)]$, which indicate the real and imaginary part of complex bands are, respectively, an even and odd function of $k$. The SLS tells that the real or imaginary part of the two complex bands is of the same magnitude and an opposite sign at any $k$. Nevertheless, the bands do not respect individually PHS$^{\dagger}$ in any phase. Rather, the complex bands are paired via PHS$^{\dagger}$: $E_{\pm}(k)=-E_{\mp}^*(-k)$. The TRS and PHS$^{\dagger}$ make the complex spectrum symmetric about the real and imaginary axis, respectively, as we show in Figs.~\ref{NSSH}(d-f). These plots also display a complex-energy gap in the trivial and topological phases, and no gap in the M{\"o}bius phase.   

When a state is simultaneous eigenstate of the Hamiltonian and TRS or SLS, the energy of the state is either purely real or zero. The states at time-reversal-invariant momenta $k=0,\pm\pi$ in the topological and trivial phase for PBC, and the edge states for an open or a special boundary conditions \cite{ritutopo2022, vyas_topological_2021} in the topological phase follow the above constraints on energy. However, such consequences of the TRS and SLS break down in the M{\"o}bius phase for $\delta \in (0, \frac{1-e^{\theta}}{1+e^{\theta}})$ at $k=\pm\pi$ for PBC as well as for the special boundary condition, when two purely imaginary energy modes emerge. The appearance of purely imaginary energy modes can be associated to the emergence of simultaneous eigenstate of the Hamiltonian and PHS$^{\dagger}$ in the intriguing M{\"o}bius phase in the parameter region between two EPs \cite{vyas_topological_2021,ritutopo2022}. Recent works~\citep {liang_topological_2013, wu_topology_2021, vyas_topological_2021, ritutopo2022} point out application of different adiabatic and non-adiabatic topological invariants for characterizing these  non-trivial topological phases with and without Hermitian counterparts. We here explore the dynamical features of these non-Hermitian topological phases with the help of the Loschmidt echo and DTOP.

\subsection{Loschmidt echo}\label{LosEcho}
The Loschmidt echo~\citep{Goussev:2012,PhysRevB.103.144305} measures the extent to which a quantum evolution can reverse upon an imperfect time reversal. It is achieved by employing slightly different forward and backward time-evolving Hamiltonians. The Loschmidt echo in Hermitian setups explored quite a lot theoretically~\citep{PhysRevLett.110.135704,PhysRevB.100.224307,PhysRevB.93.085416} as well as experimentally~\citep{Elsayed_2015,losch_exp}. The Loschmidt echo~\citep{GORIN200633,PhysRevA.106.032221,Goussev:2012} is defined as $\mathcal{L}(t)=\langle \tilde{\Psi}|e^{iH_{\mathrm{f}}t}e^{-iH_{\mathrm{i}}t}|\Psi \rangle$, where $H_\mathrm{i}$ and $H_\mathrm{f}$ are initial and final Hamiltonians with a slight difference between them. Here, $\{|\Psi\rangle,\langle\tilde{\Psi}|\}$ are mutually associated bi-orthogonal initial states. We choose $|\Psi\rangle$ as an eigenstate of $H_\mathrm{i}$, and drop the trivial phase factor from $\mathcal{L}(t)$ due to time evolution by $H_\mathrm{i}$. Thus, we simplify the above definition of the Loschmidt echo~\citep{PhysRevLett.118.015701,PhysRevB.105.184424} as
\begin{align}
\mathcal{L}(t)=\langle \tilde{\Psi}|e^{iH_\mathrm{f}t}|\Psi \rangle.\label{los}
\end{align} 
In particular, $|\Psi \rangle$ is prepared as a half-filled many-body eigenstate of $H_{\mathrm{i}}$. To perform the quench study, we change $\delta_\mathrm{i}$, $\theta_\mathrm{i}$ characterizing $H_{\mathrm{i}}$ to $\delta_\mathrm{f}$, $\theta_\mathrm{f}$ for $H_\mathrm{f}$ at $t=0$. We also keep $J=1$ everywhere. An analytical description of generic quench dynamics for our model is provided below. The numerical results of the quench dynamics and Fisher zeros (FZs) are presented in the next section.  

For a generic non-Hermitian $H_\mathrm{i}$,  $|\Psi\rangle$ as a half-filled many-body eigenstate of $H_\mathrm{i}$ occupying the lower band reads
\begin{align}
|\Psi\rangle=\displaystyle\prod_k|\psi_k^{\mathrm{i},-}\rangle=\displaystyle\prod_k\frac{1}{\sqrt{2}}\Big(
 \frac{v_\ell+w_re^{ik}}{\epsilon_k}\tilde{a}^\dagger_{k}- \tilde{b}^\dagger_{k} \Big)|0\rangle.
\end{align} 
Using $|\Psi\rangle$ in Eq.~\ref{los}, we then derive
\begin{align}\label{LE_eqn}
\mathcal{L}(t)=\displaystyle\prod_k \mathcal{G}_k(t)=\displaystyle\prod_k \langle\chi^{\mathrm{i},-}_k(0)|\psi^{\mathrm{i},-}_k(t)\rangle,
\end{align}
 where 
\begin{align}
\mathcal{G}_k(t)=\beta_k^+ e^{i\epsilon^\mathrm{f}_k t}+\beta_k^- e^{-i\epsilon^\mathrm{f}_k t}= \cos(\epsilon^\mathrm{f}_kt)-i\;\hat{d}^\mathrm{i}_k.\hat{d}^\mathrm{f}_k\sin(\epsilon^\mathrm{f}_kt),
\end{align} 
with, $\beta_k^\mu \equiv \langle\chi^{\mathrm{i},-}_k|\psi_k^{\mathrm{f},\mu}\rangle\langle\chi^{\mathrm{f},\mu}_k|\psi^{\mathrm{i},-}_k\rangle=(1-\mu\hspace{0.03cm}\hat{d}_k^{\mathrm{i}}.\hat{d}_k^\mathrm{f})/2$ for $\mu=\pm$, and $\hat{d}_k^{\mathrm{i/f}}=\frac{\vec{d_k}^{\mathrm{i/f}}}{\sqrt{\vec{d_k}^{\mathrm{i/f}}.\vec{d_k}^{\mathrm{i/f}}}}$. We here use i,f either in the subscript or in the superscript of different variables to associate them with the initial and final Hamiltonian, respectively. 

The return rate ($I(t)$) characterizes how often the evolved state after the quench comes close to the initial state. It is defined as $I(t)=-\log |\mathcal{L}(t)|/L$, which in the thermodynamic limit ($L\to \infty$) reads
\begin{align}
I(t)=-\frac{1}{2\pi}\displaystyle\oint_{k}\log |\mathcal{G}_k(t)|dk.
\end{align}
If we compare $\mathcal{L}(t)$ to the dynamical partition function for boundary state $|\Psi \rangle$ separated by $z=it$, $I(t)$ denotes the dynamical free energy for the process. The return of the evolving state $|\psi^{\mathrm{i},-}_k(t)\rangle$ after a quench to the initial state $|\psi^{\mathrm{i},-}_k(0)\rangle$ is characterized by $\mathcal{L}(t)=1$ or $I(t)=0$.  The non-analyticities in $I(t)$ indicate the DPT in the model, which occurs for $|\mathcal{L}(t)|= 0$ or $\mathcal{G}_k(t)=0$ for any $k$~\citep{PhysRevLett.110.135704}. The solutions for $|\mathcal{L}(t)|= 0$ are also known as the FZs. The FZs for $\mathcal{L}(t)$ in Eq.~\ref{LE_eqn} can be determined from $w(n,k)$ for different positive integer $n$, where
\begin{align}
w(n,k):=i\frac{\pi(2n+1)}{2\epsilon^\mathrm{f}_k}+\frac{1}{\epsilon^\mathrm{f}_k}\tanh^{-1}(\hat{d}^\mathrm{i}_k.\hat{d}^\mathrm{f}_k).
\label{fisher_zero_eq}
\end{align}
Similar to the Hermitian case, the critical times for the DPT are calculated from the imaginary intercept of the curve $w(n,k)$ over the BZ, $k\in(-\pi,\pi]$ on the parametric complex plane~\citep{PhysRevLett.110.135704,PhysRevX.11.041018,jafari_dynamical_2019}. The momentum for which the imaginary intercept happens is termed critical momentum ($k^{n}_c$), and  is found from the solution of the equation $\text{Re}[w(n,k^{n}_c)]=0$, i.e.,     
\begin{align}
\pi(n+\frac{1}{2})\text{Im}[\epsilon^\mathrm{f}_{k^{n}_c}]+\text{Re}[(\epsilon^\mathrm{f}_{k^{n}_c})^*\tanh^{-1}(\hat{d}^\mathrm{i}_{k^{n}_c}.\hat{d}^\mathrm{f}_{k^{n}_c})]=0.
\label{cirticalk}
\end{align}
For Hermitian models (i.e., $\text{Im}[\epsilon^\mathrm{f}_{k}]=0$), the above equation can be expressed as $\tanh^{-1}(\hat{d}^\mathrm{i}_{k^{n}_c}.\hat{d}^\mathrm{f}_{k^{n}_c})=0$. Consequently, the FZs emerge symmetrically in the BZ at $\pm k_c^n$. For the quenches by a non-Hermitian $H_\mathrm{f}$ in the trivial and topological phase in Secs.~\ref{sbsecA}, \ref{sbsecC}, $\text{Im}[\epsilon^\mathrm{f}_{k}] \ll \text{Re}[\epsilon^\mathrm{f}_{k}]$ as can be seen in Figs.~\ref{NSSH} (a,c). Thus, the  FZs emerge in both the positive and negative BZ like the Hermitian case. However, the contribution from $\text{Im}[\epsilon^\mathrm{f}_{k_c}]$ in Eq.~\ref{cirticalk} disrupts the  above symmetry of FZs for a non-Hermitian $H_\mathrm{f}$ in comparison for a Hermitian $H_\mathrm{f}$. Therefore, the FZs occur at different values of $k_c^{n+}$ and $k_c^{n-}$ in the BZ. The corresponding critical time ($t^{n\pm}_c$) when the DPT occurs is $\text{Im}[w(n,k^{n\pm}_c)]$, i.e.,
\begin{align}
 t_{c}^{n\pm}=\frac{\pi(n+\frac{1}{2})\text{Re}[\epsilon^\mathrm{f}_{k^{n\pm}_c}]+\text{Im}[(\epsilon^\mathrm{f}_{k^{n\pm}_c})^*\tanh^{-1}(\hat{d}^\mathrm{i}_{k^{n\pm}_c}.\hat{d}^\mathrm{f}_{k^{n\pm}_c})]}{|\epsilon^\mathrm{f}_{k^{n\pm}_c}|^2}.
\label{cri_time}
\end{align}
In the Hermitian limit (i.e., $\text{Im}[\epsilon^\mathrm{f}_{k}]=0$), the Eq.~\ref{cri_time} $t_{c}^{n\pm}=\pi(n+\frac{1}{2})/|\epsilon^\mathrm{f}_{k^{n\pm}_c}|$, which is degenerate for positive and negative critical momentum and also periodic in time. However, the contribution from second term in Eq.~\ref{cri_time}, particularly due to the real part of band energy, breaks the above degeneracy and periodicity. For $H_\mathrm{f}$ in the M{\"o}bius phase, the real and imaginary parts of $\epsilon^\mathrm{f}_{k}$ are comparable as shown in Fig.~\ref{NSSH}(b), and we need to  carefully understand the role of $\text{Im}[\epsilon^\mathrm{f}_{k}]$ in Eq.~\ref{cirticalk}. Since $\text{Im}[\epsilon^\mathrm{f}_{k}]$ and $\text{Re}[\epsilon^\mathrm{f}_{k}]$ are, respectively, an even and odd function of $k$, the relation in Eq.~\ref{cirticalk} can only be satisfied in one side (either negative or positive) of the BZ as we find in Secs.~\ref{sbsecB}, \ref{sbsecD}.

The non-analyticity in $I(t)$ indicates the vanishing overlap between evolved and initial states for some $k^{n\pm}_c$ at time $t^{n\pm}_c$, i.e., $\langle \chi^{\mathrm{i},-}_{k^{n\pm}_c}(0)|\psi^{i,-}_{k^{n\pm}_c}(t^{n\pm}_c)\rangle=0$. Such orthogonality between states emerges due to transition of a particle at momentum $k^{n\pm}_c$ from the filled lower band to the empty upper band at time $t^{n\pm}_c$. 

\subsection{Dynamical topological order parameter (DTOP)}
Due to translation symmetry within the PBC, the momentum $k$ is a good quantum number for our model, and it remains invariant under quench as we do not change the total length of the chain. The dynamical evolution of each particle is thus independent and separable from others. We can quantify the phase $\phi^G(k,t)$ of such non-adiabatic evolution of a single particle at any time $t$ using the overlap of consecutive instantaneous non-orthogonal states following Pancharatnam's description, which leads to  $\phi^{G}(k,t)=\phi^{LE}(k,t)-\phi^{dyn}(k,t)$~\citep{PhysRevB.100.224307,pancharatnam_generalized_1956}. Here, $ \phi^{LE}(k,t)$ and $\phi^{dyn}(k,t)$ denote, respectively, the Loschmidt echo phase and the dynamical phase for the evolution as given by
\begin{align}
\phi^{LE}(k,t)&=-i\log\frac{\mathcal{G}_k(t)}{|\mathcal{G}_k(t)|}, \label{pLE}\\
\phi^{dyn}(k,t)&=-i\int_0^t d\tau \;\langle \chi_k^{\mathrm{i},-}(\tau)|\frac{d}{d\tau}|\psi_k^{\mathrm{i},-}(\tau)\rangle,\label{pdyn}\\
|\psi_k^{\mathrm{i},-}(\tau)\rangle&=\frac{e^{iH_\mathrm{f}\tau}|\psi_k^{\mathrm{i},-}(0)\rangle}{\sqrt{\langle \chi_k^{\mathrm{i},-}(0)|e^{-iH^\dagger_\mathrm{f}\tau}e^{iH_\mathrm{f}\tau}|\psi_k^{\mathrm{i},-}(0)\rangle}},\label{psiE}\\
\langle\chi_k^{\mathrm{i},-}(\tau)|&=\frac{\langle\chi_k^{\mathrm{i},-}(0)|e^{-iH^\dagger_\mathrm{f}\tau}}{\sqrt{\langle \chi_k^{\mathrm{i},-}(0)|e^{-iH^\dagger_\mathrm{f}\tau}e^{iH_\mathrm{f}\tau}|\psi_k^{\mathrm{i},-}(0)\rangle}}.\label{xiE}
\end{align}
Here, $|\psi_k^{\mathrm{i},-}(\tau)\rangle$ is the instantaneous normalized ket vector at time $\tau$, and the associated bra vector is $\langle\chi_k^{\mathrm{i},-}(\tau)|$. In order to access the topological change of the aforementioned dynamic process, a dynamical topological order parameter is defined as a time-dependent winding number ($\nu_\pm(t)$) over the positive or negative momenta of the BZ~\citep{PhysRevB.93.085416}. It is given by the relation 
\begin{align}
\nu_\pm(t)=-\frac{1}{2\pi}\displaystyle\int_0^{\pm\pi} \frac{\partial\phi^G(k,t)}{\partial k}dk.
\label{dtop_eq}
\end{align}
For a fully Hermitian quench between initial and final Hermitian SSH Hamiltonian in different topological phases, the chiral symmetry of SSH chain leads to the quantization of the winding number $\nu_\pm(t)$ and enforces the condition $\phi^G(0,t \ne t_c^n)=\phi^G(\pi,t\ne t_c^n)=0$~\citep{PhysRevB.91.155127}. Consequently, for $k\in(0,\pi]$, the function $\phi^G(t):[0,2\pi]\to S^1$ represents a continuous curve on the unit circle. For quenches with a final Hamiltonian of the NSSH model, the above relation for the Pancharatnam phase remains true. We show below in Sec.~\ref{results} that $\nu_\pm(t)$ remains almost quantized for topology changing quenches by a non-Hermitian final Hamiltonian deep inside the trivial or topological phase. 

  The existence of DTOPs is generally argued due to the presence of a non-trivial topological change between the initial and final Hamiltonian~\citep{PhysRevB.100.224307, PhysRevB.91.155127, heyl_dynamical_2018}. It has been observed that quenches between topologically distinct Hamiltonians with a change in winding number $\Delta \omega$ give at least $2\Delta \omega$ topologically protected DTOPs over the BZ~\citep{PhysRevB.91.155127}. Therefore, a non-zero DTOP indicates a dynamical topological phase transition induced by quenching between topologically nonequivalent Hamiltonians. In contrast, for quenches between topologically equivalent Hamiltonians, the DTOP is zero, which indicates the sum over the change in the Pancharatnam geometric phases of all momenta, i.e., $\nu_\pm(t)$ is zero. For the NSSH model, while the geometric phase or topological invariant in the trivial or topological phase is defined for individual bands, it is only defined for the both bands together in the M{\"o}bius phase. This poses challenge to relate the change in winding number to number of DTOPs. Here, we implement different quenches by non-Hermitian Hamiltonian to address these issues regarding the appearance of DTOPs and their counts. A subsequent interpretation of these findings is the main result of our present work.
  
\section{\label{results}Results}
We now present our numerical results of the quench dynamics due to a non-Hermitian SSH final Hamiltonian to an initial state $ |\Psi \rangle$ of the Hermitian SSH model, which leads to $\mathcal{L}(t)=\langle \Psi|e^{iH_\mathrm{f}t}|\Psi \rangle$. Thus, the initial state of $H_\mathrm{i}$ can be in the trivial or  topological phase, and the parameters of $H_\mathrm{f}$ can be chosen to be in the trivial, M\"{o}bius, topological phase. Out of such six different possibilities of quench, we focus on the following four nonequivalent topology changing quenches from initial to final Hamiltonian: (1) topological to trivial,  (2) topological to M\"{o}bius, (3)  trivial to topological, and (4) trivial to M\"{o}bius.

\subsection{\label{sbsecA}Topological $\to$ Trivial} 
We start by investigating how the features of Loschmidt echo alter from a fully Hermitian quench for a non-Hermitian $H_\mathrm{f}$. So we explore a quench from an initial Hermitian topological phase with parameters, $\delta_\mathrm{i}=0.9$, $\theta_\mathrm{i}=0$, to a non-Hermitian trivial phase of parameters, $\delta_\mathrm{f}=-0.9$, $\theta_\mathrm{f}=0.4$. 
\begin{figure}[h!]
\centering
\includegraphics[scale=0.525]{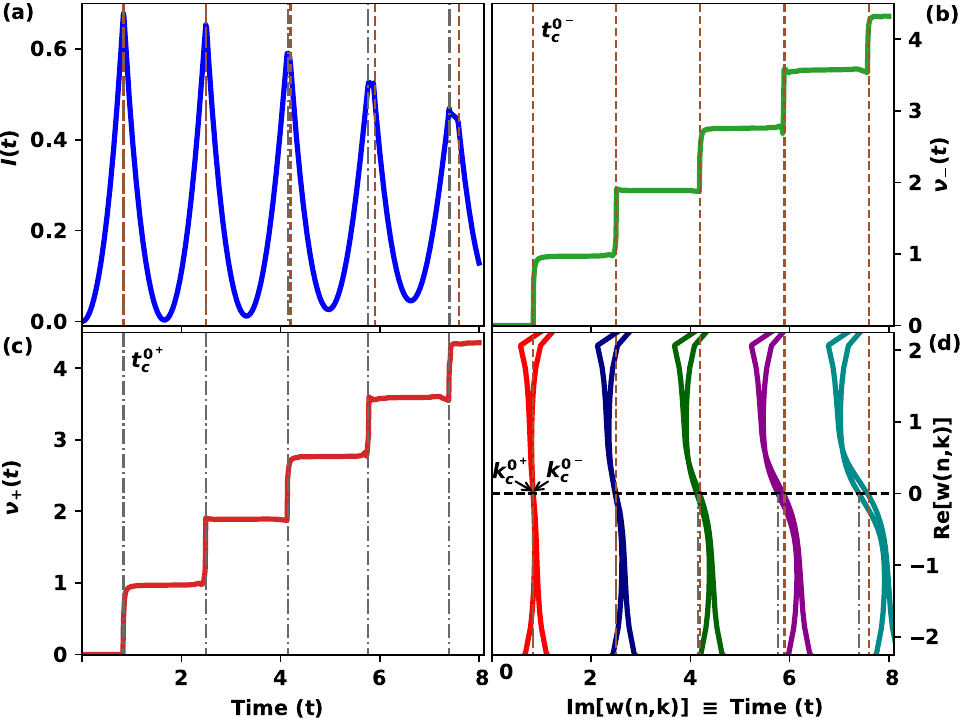}
\caption{Return rate $I(t)$ in (a), DTOP over negative momentum $\nu_-(t)$ in (b), and DTOP over positive momentum $\nu_{+}(t)$ in (c) as a function of time. Fisher zero lines in the complex plane of $w(n,k)$ over the BZ, $k\in(-\pi,\pi]$ (for n = 0,1,..4) in (d). Points of intersection of the Fisher zero lines and the dashed black line indicate critical momentum $k^{n\pm}_c$. Dashed brown (dashed-dotted grey) vertical lines through these points correspond to critical times $t_c^{n-}$($t_c^{n+}$). Initial and final Hamiltonian parameters are $\delta_\mathrm{i}=0.9, \theta_\mathrm{i}=0, \delta_\mathrm{f}=-0.9, \theta_\mathrm{f}=0.4$.}
\label{figNTT}
\end{figure}
Fig.~\ref{figNTT} displays the return rate, DTOP, and FZs for such quench dynamics. Despite the non-Hermitian complex energies involved in quench dynamics, the non-analyticities (sharp cusps) in $I(t)$ in Fig.~\ref{figNTT}(a) survive, which can be sharp as in the fully Hermitian case~\citep{sirker_boundary_2014}. Nevertheless, the periodicity of these peaks is broken for non-Hermitian quenches due to the lifting of degeneracies of the FZs, which are present in the fully Hermitian case. In Fig.~\ref{figNTT}(d), the FZs for various curves of different $n$ are shown using the cuts by the horizontal dashed line with these curves over the entire BZ, which confirms the presence of these non-analyticities at critical $k^{n\pm}_c$s.  

The imaginary values of these curves $w(n,k)$ at $k^{(n\pm)}_c$s in Fig.~\ref{figNTT}(d) are indicated by the vertical lines, and they represent the critical times $t^{n\pm}_c$ for the DPT. In Figs.~\ref{figNTT}(b,c), we show the associated DTOPs, $\nu_\pm(t)$, as a function of time for a quantitative characterization of the DPT. Both the DTOPs depict a discontinuity precisely at the critical times and remain constant between two consecutive non-analyticities in Figs.~\ref{figNTT}(b,c). The nearly quantized sharp jumps in the DTOP have been associated with a dynamical change in the topological character of the evolving state from the initial ground state with particular topological features, and the plateaus between the sharp jumps in the DTOP indicate no change in the topological nature of the evolving state. For quenches in fully Hermitian models, $k^{n+}_c=-k^{n-}_c$ and $t^{n+}_c=t^{n-}_c$. Such degeneracy in $k^{n\pm}_c$ and $t^{n\pm}_c$ is lifted due to a small but non-zero imaginary part of band energies. The maximum imaginary band energy for our particular choice of $H_\mathrm{f}$ is $\pm0.025i$, which is relatively small than real energy of order of $\pm2$. Thus, the split in $|k^{n\pm}_c|$ and the related periodicity in time is less. The FZs in Fig.~\ref{figNTT}(d) show these splits over the entire BZ for each $n$, which explains the jumps in the DTOP calculated for the negative and positive momentum in Figs.~\ref{figNTT}(b,c).

\subsection{\label{sbsecB}Topological $\to$ M{\"o}bius} 
\begin{figure}[h!]
\centering
\includegraphics[scale=0.525]{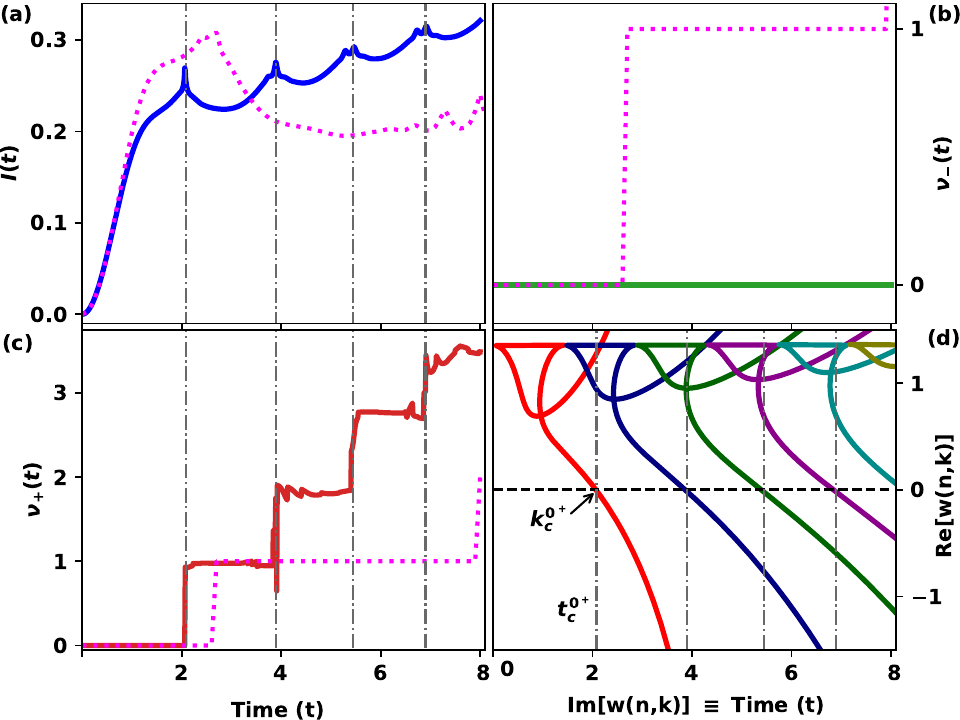}
\caption{Return rate $I(t)$ in (a), DTOP over negative momentum $\nu_-(t)$ in (b), and DTOP over positive momentum $\nu_{+}(t)$ in (c) as a function of time. Fisher zero lines in the complex plane of $w(n,k)$ over the BZ, $k\in(-\pi,\pi]$ (for n = 0,1,..4) in (d). Parameters for topological to M\"{o}bius quench are $\delta_\mathrm{i}=0.9, \theta_\mathrm{i}=0\Rightarrow\delta_\mathrm{f}=-0.1, \theta_\mathrm{f}=0.4$. Vertical dashed-dotted grey lines show critical times ($t^{n+}_c$) corresponding to positive critical momenta ($k^{n+}_c$). $I(t)$ and DTOPs in the Hermitian limit ($\theta_{\text f}=0$) are depicted by pink dotted lines.}
\label{figNTM}
\end{figure}

In this paper, we are mainly interested in studying the quench dynamics by the non-Hermitian $H_{\text f}$ in M{\"o}bius phase, which does not have a Hermitian counterpart. We here prepare the initial state $|\psi^{\mathrm{i},-}_{k}(0)\rangle$ in the topological regime of the Hermitian SSH chain (e.g., $\delta_\mathrm{i}=0.9$, $\theta_\mathrm{i}=0$), and quench it by non-Hermitian $H_{\text f}$ with parameters, $\delta_\mathrm{f}=-0.1$, $\theta_\mathrm{f}=0.4$, hosting the M{\"o}bius phase. The Fig.~\ref{figNTM}(a) shows $I(t)$ with $t$, which displays peaks indicating non-analyticities at critical time $t_c^{n+}$s. However, the height of these peaks is relatively low compared to the previous case in Fig.~\ref{figNTT}(a), and $I(t)$ increases overall with time rather than returning closer to zero.
The last trend throws light on the evolving state $|\psi^{\mathrm{i},-}_{k}(t)\rangle$, which never returns closer to $|\psi^{\mathrm{i},-}_{k}(0)\rangle$ after such a quench and moves far away from $|\psi^{\mathrm{i},-}_{k}(0)\rangle$ with progressing time. The breakdown of TRS and the emergence of PHS$^{\dagger}$ give rise to a large and purely imaginary band gap ($0.5i$ for our parameters) at the EPs in the M{\"o}bius phase, which leads to a stronger decay of the initial state in the quench process.  Interestingly, the FZ degeneracy is completely lifted in this case. The Fig.~\ref{figNTM}(d) shows FZs only in the positive momentum range of the BZ, and there is no FZ for negative momentum. As discussed earlier in Sec.~\ref{LosEcho}, this behavior stems from the  even and odd $k$-dependence of the real and imaginary part of $E_p(k)$, which values are mostly comparable in the M{\"o}bius phase between the EPs~\citep{kawabata_topological_2019,ritutopo2022}. We present the DTOPs of Eq.~\ref{dtop_eq} in Figs.~\ref{figNTM}(b,c), which show the presence (absence) of sharp jump in $\nu_+(t)~(\nu_-(t))$ with time in the  presence (absence) of FZs in the positive (negative) momentum BZ. While the topological invariant and winding number are only defined for both the complex bands together in the M{\"o}bius phase, we can draw an analogy here for the non-Hermitian quenches to the observation for Hermitian quenches \citep{PhysRevB.91.155127} by relating $2\Delta\omega$ number (which is one) of DTOP in the entire BZ for a change in winding number from the topological to M{\"o}bius phase as $\Delta\omega=0.5$. Thus, we assume the winding number for each band in the  M{\"o}bius phase being half (see Fig.~\ref{NSSH}(e)). Similar features of $I(t)$ also observed when the initial state $|\psi^{\mathrm{i},-}_{k}(0)\rangle$ is switched to $|\psi^{\mathrm{i},+}_{k}(0)\rangle$ as a filled upper band. However, the FZs instead appear only in the negative momentum of the BZ. 

In Figs.~\ref{figNTM}(a-c), we also include the plots of $I(t)$ and DTOPs by pink dotted lines for the fully Hermitian quench for such parameter regime by setting $\theta_\mathrm{f}=0$. These plots show a topological to trivial quench dynamics for fully Hermitian system because the non-Hermitian M\"{o}bius phase lies at the boundary of the Hermitian trivial phase for our parameters. The critical times for the Hermitian quench is longer than the related non-Hermitian quench due to a lower energy gap. The Figs.~\ref{figNTM} (b,c) also display the presence of DTOPs in both the positive and negative momentum of the BZ for a Hermitian quench. Thus, the non-Hermiticity induces a directional DTOP, which depends on the choice of initial ground state. Our results for DTOP in Fig.~\ref{figNTM} also confirm non-trivial topological character of the M\"{o}bius phase.

\subsection{\label{sbsecC}Trivial $\to$ Topological} 
\begin{figure}[h!]
\centering
\includegraphics[scale=0.525]{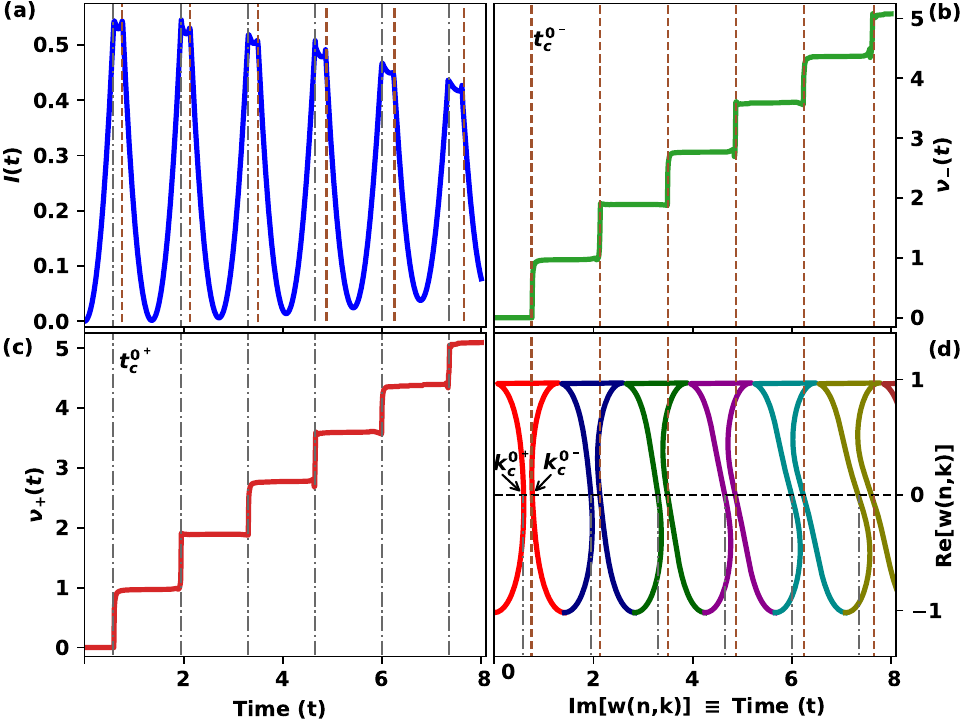}
\caption{Return rate $I(t)$ in (a), DTOP over negative momentum $\nu_-(t)$ in (b), and DTOP over positive momentum $\nu_{+}(t)$ in (c) as a function of time. Fisher zero lines in the complex plane of $w(n,k)$ over the BZ, $k\in(-\pi,\pi]$ (for n = 0,1,..4) in (d). Parameters for trivial to topological quench are $\delta_\mathrm{i}=-0.9, \theta_\mathrm{i}=0\Rightarrow\delta_\mathrm{f}=0.9, \theta_\mathrm{f}=0.4$. DTOPs in the positive and negative momentum of the BZ ($\pm$) are aligned with the critical times $t^{n\pm}_c$ (vertical lines) and the corresponding critical momenta $k^{n\pm}_c$.}
\label{figTT}
\end{figure}

Next, we consider a quench from $|\psi^{\mathrm{i},-}_{k}(0)\rangle$ of Hermitian $H_{\text i}$ in the trivial phase (e.g., $\delta_\mathrm{i}=-0.9,\;\theta_\mathrm{i}=0$) by non-Hermitian $H_{\text f}$ in the topological phase  for $\delta_\mathrm{f}=0.9,\;\theta_\mathrm{f}=0.4$. Similar to those in Subsec.~\ref{sbsecA}, $I(t)$ again features an excellent return close to the initial state at shorter times in Fig.~\ref{figTT}(a). 
Nevertheless, the sharp cusps in $I(t)$ split into two more noticeably than those in Fig.~\ref{figNTT}(a). The splittings indicate the breaking of degeneracies in the magnitude of critical positive and negative momentum as shown in Fig.~\ref{figTT}(d), and discussed after Eq.~\ref{cirticalk}. The higher splittings in Fig.~\ref{figTT}(a) in comparison to Fig.~\ref{figNTT}(a) are due to  higher values of contributing real energies (max. $2.5$) of the complex bands (Eq.~\ref{cri_time}). We observe that $\theta\to 0$ and $|\delta|\to 1$ are favorable conditions to avoid such splittings. Since the change in winding number between $H_{\text i}$ and $H_{\text f}$ is one, the FZs in Eq.~\ref{fisher_zero_eq} sweep through the imaginary axis two times as $k$ goes through the BZ in Fig.~\ref{figTT}(d). Further, the Pancharatnam geometric phase ($\phi_G(k,t)$) shows a clear jump of $2\pi$ over the negative and positive momentum range of the BZ in Fig.~\ref{fig:intro}(b), which indicates the appearance of nearly quantized DTOPs in the negative and positive momentum range in Figs.~\ref{figTT}(b,c) showing a good alignment with the corresponding critical times $t^{n\pm}_c$ and FZs in Fig. ~\ref{figTT}(d). The width of the split in the non-analyticities of $I(t)$ can be controlled by the system parameters, $\theta,\;\delta$. 

\subsection{\label{sbsecD}Trivial $\to$ M{\"o}bius}
\begin{figure}[h!]
\centering
\includegraphics[scale=0.525]{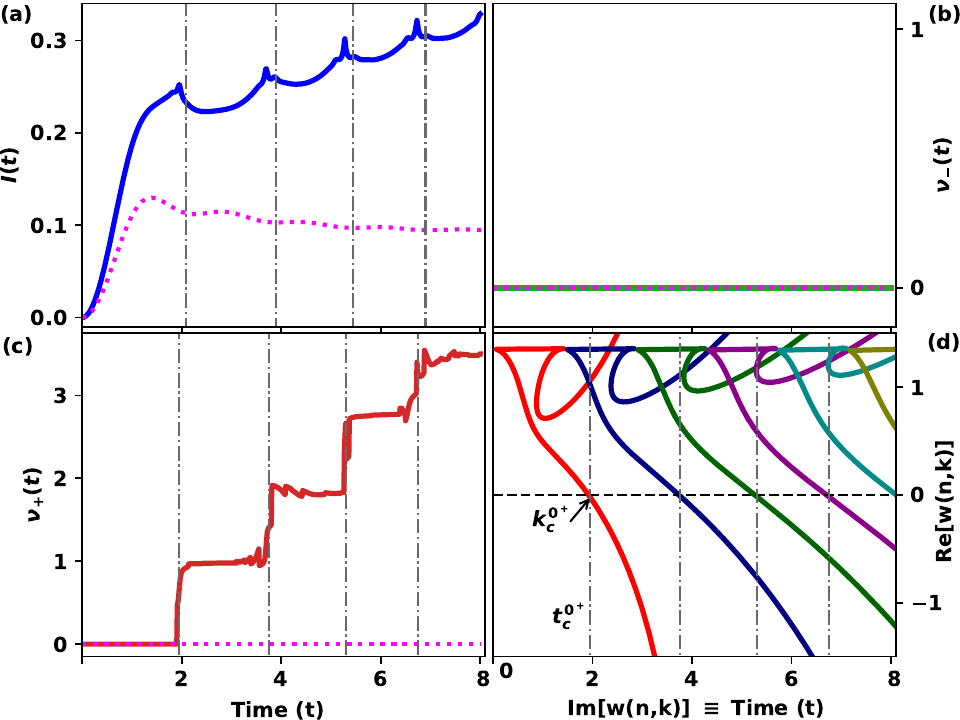}
\caption{Return rate $I(t)$ in (a), DTOP over negative momentum $\nu_-(t)$ in (b), and DTOP over positive momentum $\nu_{+}(t)$ in (c) as a function of time. Fisher zero lines in the complex plane of $w(n,k)$ over the BZ, $k\in(-\pi,\pi]$ (for n = 0,1,..4) in (d). Parameters for topological to M\"{o}bius quench are $\delta_\mathrm{i}=-0.9, \theta_\mathrm{i}=0\Rightarrow\delta_\mathrm{f}=-0.1, \theta_\mathrm{f}=0.4$. Vertical lines indicate critical times $t^{n+}_c$ for critical momenta $k^{n+}_c$ in the BZ ($k\in[0,\pi]$). $I(t)$ and DTOPs in the Hermitian limit ($\theta_{\text f}=0$) are depicted by pink dotted lines.}
\label{figTM}
\end{figure}

Since the entire M{\"o}bius phase for our choice of parameters appears in the trivial phase region of the corresponding Hermitian model, a non-trivial quench dynamics from $|\psi^{\mathrm{i},-}_{k}(0)\rangle$ of Hermitian $H_{\text i}$ in the trivial phase (e.g., $\delta_\mathrm{i}=-0.9,\;\theta_\mathrm{i}=0$) by non-Hermitian $H_{\text f}$ in the M{\"o}bius phase boundary for $\delta_\mathrm{f}=-0.1,\;\theta_\mathrm{f}=0.4$ is very significant for establishing special topological features of the M{\"o}bius phase. We remind that there is neither any non-analyticity in $I(t)$ nor a finite DTOP in the corresponding fully Hermitian quench as displayed in Fig.~\ref{figTM} by the pink dotted lines. 
The above non-Hermitian quench follows similar trends of the quench in Subsec.~\ref{sbsecB} from topological to M{\"o}bius phase. For example, $I(t)$ in Fig.~\ref{figTM}(a) shows cusps with small heights, and it moves away from the initial state with longer time scales similar to Fig.~\ref{figNTM}(a). The position of these non-analyticities around the cusps can be determined by the imaginary intercept of the FZs (in Eq.~\ref{fisher_zero_eq}) for different $n$ imprinted in Fig.~\ref{figTM}(d). Resembling the topological-M{\"o}bius quench, the Fisher zero lines cut imaginary axis only in the momentum $k\in[0,\pi]$ in Fig.~\ref{figTM}(d). The corresponding DTOPs in this range are presented in Fig.~\ref{figTM}(c), and there is no DTOP in Fig.~\ref{figTM}(b) for the negative momentum of the BZ due to the absence of FZs.

These DTOPs are reminiscent of $\phi_G(k,t)$ in Fig.~\ref{fig:intro}(a) showing $2\pi$ jump only at one side of the BZ. The other DTOP appears for negative momentum  for a quench from an initial state in the upper energy band due to the symmetry constraints on the complex energy bands' real and imaginary parts as discussed in Subsec.~\ref{sbsecB}. Comparing the plots of $I(t)$ and DTOP between the Hermitian and non-Hermitian quenches in Fig.~\ref{figTM}, we argue that non-Hermiticity gives rise to the M\"{o}bius phase with exciting dynamical topological features in the non-equilibrium setups due to spontaneous breaking of TRS and emergence of PHS$^{\dagger}$~\citep{ritutopo2022,kawabata_topological_2019}. It should be further noted that there is no significant difference obtained in $I(t)$ and DTOPs between the topological-M{\"o}bius quench and trivial-M{\"o}bius quench. The last is due to the PBC employed in these studies, which reminds us of the fully Hermitian case where a symmetric trivial-topological and a topological-trivial quench are not differentiable for PBC~\citep{PhysRevB.97.064304}.

\section{\label{conclusion}Summary and Outlook}
In this work, we have shown new dynamical properties of the recently discovered composite M{\"o}bius phase of sublattice symmetric non-Hermitian models to confirm the unique topological character of the phase. One of the main findings is the appearance of DTOP in quench dynamics by the non-Hermitian SSH Hamiltonian in the M{\"o}bius phase indicating topological differences of the M{\"o}bius phase from the other two phases (namely, trivial and topological phase) of the non-Hermitian SSH model. Moreover, the DTOP appears only at one side of the BZ for any particular choice of the initial state, which is a dynamical signature of different symmetry constraints on the real and imaginary part of the complex bands  in the M{\"o}bius phase. The above features of DTOP are significantly different for the other two non-Hermitian topological phases (trivial and non-trivial) with Hermitian counterparts, where the presence of a large complex band gap ensures two DTOPs in the BZ for quenches between these distinct topological phases differed by one winding number. The experiments on the dynamical studies of these new non-Hermitian topological phases could lead to a deeper connection between these topological states' robustness~\cite {10.1126/science.aay1064}.

We have employed the time-evolution of bi-orthogonal states given in Eqs.~\ref{psiE}-\ref{xiE} following the Schr{\"o}dinger equation and by normalizing it at each instant. This is an approximate scheme for non-Hermitian systems. Nevertheless, other schemes, such as, the metric method~\citep{Mostafazadeh2018,Mostafazadeh2020} can be tested for such dynamical studies. We believe that the position of non-analyticities of $I(t)$ at the FZs would not change for a different scheme. Further, while the conventional quantum phase transitions are accompanied by an emergence of a local order parameter due to spontaneous symmetry breaking in one phase, no such local order parameter or spontaneous symmetry breaking is associated with topological phase transition in equilibrium \cite{asboth_su-schrieffer-heeger_2016}. The topological phase transitions in Hermitian systems are instead classified by topological invariants. However, the topological changes in non-Hermitian systems are also in some cases accompanied by spontaneous symmetry breaking although without emergence of a local order parameter.  The signatures of dynamical quantum phase transitions, e.g., the non-analyticities of $I(t)$, seem similar to those for dynamical topological phase transitions. Thus, both these dynamical phase transitions have been unified in out-of-equilibrium systems. 

In our earlier work \citep{ritutopo2022}, we have extended the research on sublattice symmetric non-Hermitian SSH models by going beyond bi-partite models to tripartite and quadripartite models with three and four sublattice sites per unit cell, respectively. Such extension generates various non-Hermitian topological insulating and metallic composite phases with non-trivial topology like the Penrose triangle. It would be exciting to explore quench dynamics by final Hamiltonian in such composite phases to understand the dynamical properties of such topological phases. While there are many experimental realizations of $\mathcal{PT}$ symmetric non-Hermitian models in engineered photonic, cold atomic, phononic, and electrical systems, the nonreciprocal hopping for sublattice symmetric non-Hermitian models is only recently realized \cite{liu2023experimental, PhysRevB.107.085426}. Thus, these non-Hermitian models can be explored to experimentally probe $I(t)$ and DTOPs following the implementation in the Hermitian settings~\citep{PhysRevLett.110.135704, Elsayed_2015,losch_exp,heyl_dynamical_2018,PhysRevB.105.094514,PhysRevLett.118.015701,PhysRevB.105.184424}.

\section{Acknowledgments}
We thank Kiran Estake for many useful discussions.

%\bibliography{My_Library}
\bibliography{AnoLosEcho.bib}
\end{document}